\documentclass[letterpaper, 10 pt, conference]{ieeeconf}  

\IEEEoverridecommandlockouts 

\usepackage{amssymb}
\usepackage{mathrsfs,amsmath}
\usepackage{graphicx}
\usepackage{amsmath}
\usepackage{bbm}
\usepackage{dsfont}
\usepackage{listing}
\usepackage{hyperref}
\usepackage{color}
\usepackage[usenames,dvipsnames]{xcolor}
\usepackage[linesnumbered,ruled]{algorithm2e}

\newtheorem{lemma}{Lemma}

\newtheorem{definition}{Definition}

\newif\ifmargincomments 
\margincommentstrue 

\ifmargincomments
\newcommand{\mpmargin}[2]{{\color{cyan}#1}\marginpar{\color{cyan}\raggedright\footnotesize [MP]:
#2}}

\newcommand{\rimargin}[2]{{\color{blue}#1}\marginpar{\color{blue}\raggedright\footnotesize [RI]:#2}}

\else
\newcommand{\mpmargin}[2]{#1}

\newcommand{\rimargin}[2]{#1}

\fi

\newif\ifarxiv

\arxivfalse

\graphicspath{ {fig/} }

\title{\LARGE \bf
Stochastic Model Predictive Control \\ for Autonomous Mobility on Demand
}

\author{Matthew Tsao$^{1}$, Ramon Iglesias$^{2}$, and Marco Pavone$^{3}$
\thanks{$^{1}$Matthew Tsao is with the Department of Electrical Engineering,
        Stanford University, 496 Lomita Mall, Stanford, CA 94102, USA
        {\tt\small mwtsao@stanford.edu}}%
\thanks{$^{2}$Ramon Iglesias is with the Department of Civil Engineering,
        Stanford University, 496 Lomita Mall, Stanford, CA 94102, USA
        {\tt\small rdit@stanford.edu}}%
\thanks{$^{3}$Marco Pavone is with the Department of Aeronautics and Astronautics,
        Stanford University, 496 Lomita Mall, Stanford, CA 94102, USA
        {\tt\small pavone@stanford.edu}}%
\thanks{This research was supported by the National Science Foundation under CAREER Award CMMI-1454737 and the Toyota Research Institute (TRI). This article solely reflects the opinions and conclusions of its authors and not NSF, TRI, or any other entity.}
}

\hypersetup{draft}
\begin{document}

\maketitle
\thispagestyle{empty}
\pagestyle{empty}

\begin{abstract}
This paper presents a stochastic, model predictive control (MPC) algorithm that leverages short-term probabilistic forecasts for dispatching and rebalancing Autonomous Mobility-on-Demand systems (AMoD), i.e. fleets of self-driving vehicles. We first present the core stochastic optimization problem in terms of a time-expanded network flow model. Then, to ameliorate its tractability, we present two key relaxations. First, we replace the original stochastic problem with a Sample Average Approximation, and provide its performance guarantees. Second, we divide the controller into two submodules. The first submodule assigns vehicles to existing customers and the second redistributes vacant vehicles throughout the city. This enables the problem to be solved as two totally unimodular linear programs, allowing the controller to scale to large problem sizes. Finally, we test the proposed algorithm in two scenarios based on real data and show that it outperforms prior state-of-the-art algorithms. In particular, in a simulation using customer data from the ridesharing company DiDi Chuxing, the algorithm presented here exhibits a 62.3 percent reduction in customer waiting time compared to state of the art non-stochastic algorithms.
\end{abstract}

\IEEEpeerreviewmaketitle

\section{Introduction}
The last decade has witnessed a rapid transformation in urban mobility. On the one hand, Mobility-on-Demand (MoD) services like ridesharing (e.g. Uber and Lyft) and carsharing (e.g. Zipcar, Car2Go) have become ubiquitous due to the convenience and flexibility of their services. On the other hand, the advent of self-driving vehicles promises to further revolutionize urban transportation. Indeed, some expect that the junction of these two paradigms, Autonomous Mobility-on-Demand (AMoD) 
will have such a profound impact that it will dramatically reduce personal vehicle ownership \cite{Neil}.


In particular, AMoD systems present a unique opportunity to address the widespread problem of vehicle \textit{imbalance}: any uncontrolled MoD system will inevitably accumulate vehicles in some areas and deplete others \cite{David2012}, hampering the quality of service. Unlike existing MoD systems, an AMoD operator can order empty, self-driving vehicles to rebalance themselves.

Accordingly, this opportunity has spurred the development of controllers that attempt to optimally rebalance AMoD systems in real time. However, as we discuss in the literature review, most of the existing controllers either ignore future demand, assume deterministic future demand, or do not scale to large systems. In particular, while travel demand follows relatively predictable patterns, it is subject to significant uncertainties due to externalities such as, e.g., weather and traffic. Successful AMoD systems must cope with these uncertainties. 
Thus, the goal of this paper is to propose a stochastic model-predictive control approach for vehicle rebalancing that leverages short-term travel demand forecasts while considering their uncertainty. 

{\emph{Literature Review.}}
To keep this paper concise, we limit our review to work that specifically addresses AMoD systems, although similar ideas can be found in the MoD literature.
We categorize prior work in real-time control of AMoD systems in two broad classes: i) reactive control methods that do not make assumptions about future demand and ii) Model Predictive Control (MPC) algorithms that are able to leverage signals about future demand. 
Reactive, time-invariant methods span from simple bipartite matching, to control methods based on fluidic frameworks. A good comparison of different reactive controllers can be found in \cite{ZhangRossiEtAl2016b} and \cite{HoerlRuchEtAl2018}, where, notably, both studies show that the controller first proposed in \cite{PavoneSmithEtAl2012} performs competitively across tests.
However, these controllers do not provide a natural way to leverage travel demand forecasts.

In contrast, time-varying MPC algorithms, such as those proposed in \cite{ZhangRossiEtAl2016b,IglesiasRossiEtAl2018,MiaoHanEtAl2017}, provide a natural way to leverage travel forecasts. However, \cite{ZhangRossiEtAl2016b} suffers from computational complexity as the fleet size grows, and \cite{ZhangRossiEtAl2016b,IglesiasRossiEtAl2018} do not account for uncertainty on the forecasts. While the authors of \cite{IglesiasRossiEtAl2018} show impressive results in their experiments, it can be shown that the difference between the stochastic optimum and the certainty equivalent one can be arbitrarily large. 
To address stochasticity of demand, \cite{MiaoHanEtAl2017} proposes a distributionally robust approach leveraging semidefinite programming. However, their model makes a restrictive Markovian assumption that exchanges fidelity for tractability. Moreover, the authors do not address how to recover integer rebalancing tasks from the fractional strategy provided by the controller.

To the best of our knowledge, there is no existing AMoD controller that i) exploits travel demand forecasts while considering its stochasticity, ii) produces actionable integer solutions for real-time control of AMoD systems, and  iii) scales to large AMoD systems. 

{\emph{Statement of Contributions.}} The contributions of this paper are threefold. 
First, we develop a stochastic MPC algorithm that leverages travel demand forecasts and their uncertainties to assign and reposition empty, self-driving vehicles in an AMoD system. 
Second, we provide high probability bounds on the suboptimality of the proposed algorithm when competing against an oracle controller which knows the true distribution of customer demand. 
Third, we demonstrate through experiments that the proposed algorithm outperforms the aforementioned deterministic counterparts when the demand distribution has significant variance. In particular, on the same DiDi Chuxing dataset, our controller yields a 62.3 percent reduction in customer waiting time compared to the work presented in \cite{IglesiasRossiEtAl2018}.

{\emph{Organization.}} The remainder of the paper is organized as follows. We introduce the AMoD rebalancing problem in Section~\ref{sec:problem-formulation} and we formulate it as an explicit stochastic integer program using a Sample Average Approximation (SAA) approach in Section~\ref{sec: SAA-ILP}. In Section~\ref{sec: convex-relax} we discuss approximation algorithms to rapidly solve such an integer program, while in Section~\ref{sec: MPC} we leverage the presented results to design a stochastic MPC scheme for AMoD systems. In Section~\ref{sec:numerical-experiments} we compare the proposed MPC scheme against state-of-the-art algorithms using numerical simulations. Section~\ref{sec:conclusions} concludes the paper with a brief discussion and remarks on future research directions.
 

\section{Model and Problem Formulation}\label{sec:problem-formulation}
In this section, we first present a stochastic, time-varying network flow model for AMoD systems that will serve as the basis for our control algorithms. Unlike in \cite{IglesiasRossiEtAl2018}, the model does not assume perfect information about the future, instead it assumes that customer travel demand follows an underlying distribution, which we may estimate from historical and recent data. Then, we present the optimization problem of interest: how to minimize vehicle movements while satisfying as much travel demand as possible. Finally, we end with a discussion on the merits and challenges of the model and problem formulation.

\subsection{Model}
Let $G = (\mathcal{V},\mathcal{E})$ be a weighted graph representing a road network, where $\mathcal{V}$ is the set of discrete regions (also referred to as \textit{stations}), and the directed edges $\mathcal{E}$ represent the shortest routes between pairs of stations. We consider $G$ to be fully connected so there is a path between any pair of regions. Accordingly, let $n = |\mathcal{V}|$ denote the number of stations. We represent time in discrete intervals of fixed size $\Delta t$.  The time it takes for a vehicle to travel from station $i$ to station $j$, denoted $\tau_{ij}$, is an integer multiple of $\Delta t$ for all pairs $i,j \in \mathcal{V}$. 

At time $t$, we consider a planning horizon $\mathcal{T}$ of $T$ consecutive time intervals, i.e. $\mathcal{T} = [t+1,t+2,...,t+T]$. For notational convenience and without loss of generality, we will always assume that the beginning of the planning horizon is at time $t = 0$.  For each time interval in $\mathcal{T}$, $\lambda_{ijt}$ represents the number of future passengers that want to go from station $i$ to station $j$ at time interval $t$. However, the travel demand is a random process. Thus, we assume that the travel demand $\Lambda = \{\lambda_{ijt}\}_{i,j \in \mathcal{V}, t \in \mathcal{T}}$ within the time window $\mathcal{T}$ is characterized by a probability distribution $P$. Additionally, $\lambda_{ij0}$ denotes the number of outstanding passengers who have already issued a request to travel from $i$ to $j$ some time in the past but have not yet been serviced. Note that it is safe to assume that $\lambda_{ij0}$ is always known (since keeping track of waiting customers is relatively trivial) and, therefore, deterministic.

Within the same time window, there are $m$ self-driving vehicles which are either idling, serving a customer, or executing a rebalancing task. Thus, the availability of these vehicles is location and time-dependent. Specifically, $a_i$ is the number of idle vehicles at the beginning of the time window at station $i$, and $v_{it}$ the number of vehicles which are currently busy, but will finish their current task and become available at time $t$ at station $i$. Thus, the total number of available vehicles in the system as a function of location and time is given by

\begin{align*}
	s_{it} := \left\{
	\begin{tabular}{cc}
		$a_i + v_{it}$ & if $t = 1$\,, \\
		$v_{it}$ & if $t > 1$\,.
	\end{tabular}
	\right.
\end{align*}

Vehicle movements are captured by $x$, i.e., $x_{ijt}$ is the number of cars, rebalancing or serving customers, which are departing from $i$ at time $t$ and traveling to $j$. Note that vehicles must satisfy flow conservation, such that, the number of vehicles arriving at a station at a particular time equals the number of departing vehicles. Formally:

\begin{equation}\label{eq:flow-conservation}
	\sum_{j = 1}^n  x_{ijt} = s_{it} + \sum_{j = 1}^n x_{ji(t - \tau_{ji})}\,, \forall i \in \mathcal{V}, t \in \mathcal{T}\,.
\end{equation}

Finally, $w$ captures outstanding customers, such that $w_{ijt}$ is the number of outstanding customers who waited until time $t$ to be transported from station $i$ to station $j$. All outstanding customers must be served within the planning horizon:

\begin{equation}\label{eq:waiter-conservation}
	\sum_{t = 0}^{T} w_{ijt} = \lambda_{ij0} \; \forall i,j \in \mathcal{V}\,.
\end{equation}

\subsection{Problem Formulation}
Our objective is to minimize a combination of i) the operational cost based on vehicle movement, ii) the waiting time for outstanding customers and iii) the expected number of customers who upon arrival do not find an available vehicle in their region.
Given $(\zeta)_+ := \max \{0,\zeta\}$ and a vehicle availability state $\{s_{it}\}_{i \in \mathcal{V}, t\in \mathcal{T}}$, the goal is to solve the following optimization problem:

{\small
\begin{align}\label{eq:original-problem}
& \underset{x,w}{\text{min.}} 
  & & c_x^T x + c_w^T w + \mathbb{E}_P \left[ \sum_{ijt} c_{\lambda,ijt}(\lambda_{ijt}+ w_{ijt} - x_{ijt})_+\right]\,,\\
  & \text{s.t} && \text{\eqref{eq:flow-conservation}, \eqref{eq:waiter-conservation}}\,.\nonumber \\
  & & & w,x \in \mathbb{N}^{n^2T} \nonumber
\end{align}
}
The first term in the objective, where $c_x := \{ c_{x,ijt} \}_{i,j \in \mathcal{V}, t \in \mathcal{T}}$, is the operational cost, i.e. the cost of operating the fleet (including, e.g., fuel, maintenance, depreciation) in proportion to total distance traveled. Similarly, the second term $c_w^T w$ penalizes customer waiting times by a cost vector $c_w$, where $c_{w,ijt}$ is the cost of making an outstanding customer wanting to travel between stations $i$ and $j$ wait until time interval $t$ to be served. The last term penalizes the expected mismatch between customer demand and the vehicle supply, that is $c_{\lambda,ijt}$ is the cost of not being able to serve a customer wanting to travel between $i$ and $j$ at time $t$. Finally, in addition to the previously mentioned constraints, $x$ and $w$ must be positive integers since fractional vehicles and customers are non-physical.

\subsection{Discussion}


There are two key challenges in solving \eqref{eq:original-problem}. First, $P$ is a time varying high dimensional probability distribution which is generally not known. Hence, one cannot evaluate the objective function explicitly. Secondly, due to the integer constraints on $x,w$, \eqref{eq:original-problem} is an instance of integer programming which is NP-hard, such that no polynomial time algorithms exist and the problem remains computationally intractable for large inputs.

In the following sections, we present a series of relaxations that allow us to efficiently obtain solutions to a surrogate problem that approximates \eqref{eq:original-problem}. Specifically, to address the unknown distribution in the objective function, we fit a conditional generative model on historical data to predict future demand given recent realizations of demand. To address the computational complexity of integer programming, we perform several relaxations to arrive at a linear programming surrogate problem. Finally, we present bounds on the optimality gap induced by making these relaxations. 
\section{Sample Average Approximation Techniques}\label{sec: SAA-ILP}

Since $P$ is an unknown, time varying distribution, we cannot explicitly evaluate the objective in \eqref{eq:original-problem}. To address this issue, we present a SAA problem whose objective function approximates the objective of \eqref{eq:original-problem} in section \ref{subsec: SAA_prob}. In section \ref{subsec: SAA_bound} we give sufficient conditions under which the solution to the SAA problem from \ref{subsec: SAA_prob} is near optimal for the original problem. We address the trade-off between solution accuracy and problem complexity in section \ref{subsec: SAA_scale}. \\

\subsection{Sample Average Approximation for AMoD control}\label{subsec: SAA_prob}
Despite not knowing $P$, nor being able to sample from it, we have historical data from $P$  that we use to train a conditional generative model $\widehat{P}$ to mimic the behavior of $P$. With a generative model in hand, one can consider solving \eqref{eq:original-problem} with $\widehat{P}$ instead of $P$.

However, in many cases solving a stochastic optimization problem \textit{exactly} is not possible if the underlying distribution does not have a computationally tractable form. Many popular probabilistic generative models, such as Bayesian networks and Bayesian neural networks fall into this category. To overcome this issue, we can sample from the generative model and replace expectations with Monte Carlo estimates to get \textit{approximate} solutions, a method commonly referred to Sample Average Approximation (SAA) \cite{Homem-de-MelloBayraksan2014,BirgeLouveaux2011}. To this end we generate $K$ samples $\{\{\lambda_{ijt}^k\}_{i,j \in [n], t \in [T]}\}_{k=1}^K \overset{\text{i.i.d.}}{\sim} \widehat{P}$ and approximate expectations under $\widehat{P}$ with Monte Carlo estimates, i.e.

{\small
\begin{equation*}
	 \mathbb{E}_{\widehat{P}} \left[ \sum_{ijt} (\lambda_{ijt}+ w_{ijt} - x_{ijt})_+\right] \approx \frac{1}{K} \sum_{k=1}^K  \sum_{ijt} (\lambda_{ijt}^k + w_{ijt} - x_{ijt})_+\,.
\end{equation*}
}

Using this approximation, we consider the following SAA surrogate problem:

{\small
\begin{align}\label{monte_carlo_bundled}
\underset{\{u^k\}_{k},x,w}{\text{min }} & c_x^T x + c_w^T w +\frac{c_\lambda}{K} \sum_{k=1}^K  \sum_{ijt} u_{ijt}^k\, \\
\text{ s.t.} & \sum_{t \in \mathcal{T}} w_{ijt} = \lambda_{ij0}\,\quad \forall i,j \in [n]\, \nonumber \\
& \sum_{j = 1}^n  x_{ijt} - x_{ji(t - \tau_{ji})} = s_{it}\,\quad \forall i \in [n], t \in \mathcal{T}\, \nonumber \\
& u_{ijt}^k \geq 0\,\quad \forall k \in [K], i,j \in [n],t \in \mathcal{T}\, \nonumber\\
& u_{ijt}^k \geq \lambda_{ijt}^k + w_{ijt} -x_{ijt} \, \forall k \in [K], i,j \in [n],t \in \mathcal{T}\, \nonumber \\
& \{u^k\}_{k=1}^K, x,w \in \mathbb{N}^{n^2 T} \,\quad \forall k \in [K], i,j \in [n],t \in \mathcal{T}\,, \nonumber
\end{align}
}
where, in addition to the Monte Carlo estimate, we include a series of inequalities to make the objective function linear. Specifically, minimizing $(x)_+$ is equivalent to minimizing $u$ with the constraints $u \geq 0, u \geq x$. The surrogate SAA problem  \eqref{monte_carlo_bundled} is directly solvable by off-the-shelf mixed integer linear programming (MILP) solvers.\\

\subsection{Oracle inequality performance guarantees for SAA}\label{subsec: SAA_bound}

Sample Average Approximation is not guaranteed in general to provide asymptotically optimal solutions to the population problem as the number of samples goes to infinity. While the objective of an SAA problem converges pointwise to the population objective, if the convergence is not uniform, SAA may return solutions that do not converge to the optimal population value even as the number of samples goes to infinity. In this section, we compare the quality of the solutions to \eqref{eq:original-problem} and \eqref{monte_carlo_bundled} when evaluated by the objective in \eqref{eq:original-problem}. Specifically, we present a result stating that if $\widehat{P}$ is close to $P$ in an appropriate sense and we use enough samples for the SAA in \eqref{monte_carlo_bundled}, then the obtained solution is with high probability, provably near optimal for the original problem in \eqref{eq:original-problem} that we would have solved had we known $P$. Such a result is called an oracle inequality. Using the notation


{\small
\begin{align*}
&F(x,w) := c_\lambda \mathbb{E}_P \left[ \sum_{ijt} (\lambda_{ijt} + w^*_{ijt} - x^*_{ijt})_+ \right] \text{ and } \\
&\widehat{F}_K(x,w) := \frac{c_\lambda}{K} \sum_{k=1}^K \left[ \sum_{ijt} (\lambda^k_{ijt} + \widehat{w}_{ijt} - \widehat{x}_{ijt})_+ \right],
\end{align*}
}

the difference between the objectives in \eqref{eq:original-problem} and \eqref{monte_carlo_bundled} is $F(x,w) - \widehat{F}_K(x,w)$. Consider the following lemma: \\ 
\begin{lemma}[$||\cdot||_\infty$-continuity of function minima]\label{triangle} Let $f,g : \mathcal{X} \rightarrow \mathbb{R}$ denote two real valued functions that have finite global minima, i.e., both $x_f := \arg\min_{x \in \mathcal{X}} f(x)$ and $x_g := \arg\min_{x \in \mathcal{X}} g(x)$ exist. Then,

{\small
\begin{align*}
f(x_g) &\leq f(x_f) + 2 \sup_{x \in \mathcal{X}} |f(x) - g(x)|.\\
\end{align*}
}
\end{lemma}

\noindent See section \ref{app: triangle_proof} for a proof.

Applying this idea to the AMoD setting, let $(x^*,w^*)$ be a solution to \eqref{eq:original-problem}, and $(\widehat{x},\widehat{w})$ a solution to \eqref{monte_carlo_bundled}. If $\max_{x, w} |F(x,w) - \widehat{F}_K(x,w)| < \epsilon$ is small, then $(\widehat{x}, \widehat{w})$ will be at most $2\epsilon$ worse than $(x^*,w^*)$ when evaluated by $F$. It is then of interest to understand the conditions for which $\widehat{F}_K$ will be uniformly close to $F$. Since $\widehat{F}_K$ is a random object, its error in estimating $F$ has two contributors: stochastic error and model error. Specifically, the stochastic error is due to the error induced by estimating expectations under $\widehat{P}$ using SAA, and the model error is the error incurred when estimating the true distribution $P$ using $\widehat{P}$. For the analysis, we will need the following definition. \\

\begin{definition} \textit{Sub-exponential Random Variables} \\
A random vector $X \in \mathbb{R}^d$ is sub-exponential with parameters $\sigma^2, b < \infty$ if, for any $v \in \mathbb{R}^d$ satisfying $||v||_2 \leq b^{-1}$, the following inequality holds:
{\small
\begin{align*}
\log \mathbb{E} \left[ e^{v^T(X - \mathbb{E}X)} \right] &\leq \frac{||v||_2^2 \sigma^2}{2}. \\
\end{align*}
}
\end{definition}
Intuitively, a random variable is sub-exponential if its tails decay at least as fast as that of an exponential random variable. \\
\begin{lemma}[Uniform Convergence for SAA]\label{bound} Let $P$ be the true distribution of customer demand, $\widehat{P}$ be the distribution of predicted customer demand and let $P_{ijt}, \widehat{P}_{ijt}$ be the distribution of $\lambda_{ijt}$ under $P,\widehat{P}$ respectively. Assuming that $\lambda \sim \widehat{P}$ is $(\sigma^2,b)$ sub-exponential, then for any $\delta > 0$, with probability $1-\delta$, the following holds:

\begin{small}
\begin{align}\label{subopt}
& \max_{x,w} |F(x,w) - \widehat{F}_K(x,w)|\\
&\leq \underbrace{\frac{2 \sigma}{\sqrt{K}} \sqrt{n^2 T\log(m) + \log \frac{1}{\sqrt{\delta}}}}_{\text{Stochastic Error}} + \underbrace{ ||\chi(\widehat{P}||P)||_2 \sqrt{\text{Var}_P(||\lambda||_2)} }_{\text{Model Error}} \nonumber.
\end{align}
\end{small}

where $\chi(\widehat{P}||P) \in \mathbb{R}^{n^2T}_{+}$, $\chi(\widehat{P}||P)_{ijt} = \chi(\widehat{P}_{ijt}||P_{ijt})$ and $\chi^2(\cdot||\cdot)$ represents the $\chi^2$-divergence between probability distributions which is non-negative and zero if and only if its arguments are equal. \\

\noindent See section \ref{app: bound_proof} for a proof.

Note that the assumption of sub-exponential $\lambda$ is not very restrictive. Indeed, many common distributions including gaussian, Poisson, chi-squared, exponential, geometric, and any bounded random variables are all sub-exponential \cite{vershynin2018}. If we denote the solution to \eqref{eq:original-problem} as $(x^*,w^*)$ and the solution to \eqref{monte_carlo_bundled} as $(\widehat{x}, \widehat{w})$, then applying lemmas \ref{triangle} and \ref{bound}, the following happens with probability at least $1-\delta$. 

{\small
\begin{align*}
& \frac{1}{2} \left( F(\widehat{x}, \widehat{w}) - F(x^*, w^*) \right) \\
&\leq \underbrace{\frac{2 \sigma}{\sqrt{K}} \sqrt{n^2 T\log(m) + \log \frac{1}{\sqrt{\delta}}}}_{\text{Stochastic Error}} + \underbrace{||\chi(\widehat{P}||P)||_2 \sqrt{\text{Var}_P(||\lambda||_2)} }_{\text{Model Error}}. \\
\end{align*}
}

This result implies that, for a desired accuracy $\epsilon > 0$, if we fit a generative model $\widehat{P}$ satisfying $||\chi(\widehat{P}||P)||_2 \leq 0.25 \epsilon \text{Var}(||\lambda||_2)^{-1/2}$ and we use at least $K_\epsilon \geq 64 \sigma^2 \epsilon^{-2} \left( n^2 T \log (m) - 0.5 \log \delta \right)$ samples for the SAA in \eqref{monte_carlo_bundled}, then the solution to \eqref{monte_carlo_bundled} will be at most $\epsilon$ worse than the optimal solution to \eqref{eq:original-problem} with known $P$ . 
\end{lemma}

\subsection{Computational Complexity}\label{subsec: SAA_scale}
As shown in lemma \ref{bound}, the sampling error of \eqref{monte_carlo_bundled} is $O(K^{-1/2})$, where $K$ is the number of samples used to form the SAA objective. On the other hand, the computational complexity of \eqref{monte_carlo_bundled} is an increasing function of $K$, so in this section we discuss how the problem size of \eqref{monte_carlo_bundled} depends on $K$. A naive implementation of \eqref{monte_carlo_bundled} would allocate $Kn^2 T$ decision variables for $\{u^k\}_{k=1}^K$, and a linear dependence on $K$ which would lead to scalability issues since integer programming is NP hard in the worst case. However, note that in an optimal solution, we will have $u_{ijt}^k = (\lambda_{ijt}^k + w_{ijt} - x_{ijt})_+$. Thus if for some $k,l$ we have $\lambda_{ijt}^k = \lambda_{ijt}^l$, then the optimal solution has $u_{ijt}^k = u_{ijt}^l$. In this case, solving \eqref{monte_carlo_bundled} with the additional constraint of $u_{ijt}^k = u_{ijt}^l$ will still yield the same optimal value while reducing the number of decision variables by one. Therefore, for each trip type $(i,j,t)$, instead of needing $K$ decision variables $\{u_{ijt}^k\}_{k=1}^K$, we only need $c$ decision variables, where $c$ is the number of unique values in the set $\{\lambda_{ijt}^k\}_{k=1}^K$. The following lemma demonstrates the reduction in complexity achievable by this variable elimination procedure. \\

\begin{lemma}[SAA Problem Size for Subexponential Demand] \label{num_vars}
Assume that $\lambda \sim \widehat{P}$ is sub-exponential with parameters $\sigma^2,b$. For any $\delta > 0$, with probability at least $1-\delta$, the number of distinct realizations of the customer demand is no more than  $O \left( n^2 T \min\left(\log \frac{K n^2 T}{\delta}, K\right) \right)$. Thus, as long as $n^2T$ is not exponentially larger than $K$, a variable elimination procedure ensures that the number of decision variables scales as $O(\log K)$, as opposed to the linear scaling $O(K)$ that the naive implementation would lead one to believe. \\
\end{lemma}
\noindent See section \ref{app: num_vars_proof} for a proof.

Thus with high probability the number of decision variables will be logarithmic in $K$, which is an exponential improvement over the linear dependence that the naive implementation proposes. This is especially important since using large $K$ gives an objective function with less variance.

\section{Scalable Integer Solutions via Totally Unimodular Linear Relaxations} \label{sec: convex-relax}

Recall from \eqref{subopt} that increasing the number of samples $K$ used for Monte Carlo reduces the standard deviation of the random objective in \eqref{monte_carlo_bundled}, thereby increasing the quality of the algorithm's output. While we showed that the number of decision variables is only logarithmic in the sample size $K$, the problem is still NP-hard. Thus, increasing the number of samples used in \eqref{monte_carlo_bundled} may not be tractable in large scale settings. In this section, we propose a modified algorithm that solves a convex relaxation of \eqref{monte_carlo_bundled}, which is scalable to large problem sizes. \\
\indent Our relaxation separately addresses the tasks of servicing existing customers and rebalancing vacant vehicles that are jointly solved in \eqref{monte_carlo_bundled}. Note that information about future customers can affect scheduling of waiting customers and vice versa in the optimal solution. In such a situation, servicing existing customers and rebalancing vacant vehicles with two separate algorithms prevents the sharing of information and can lead to suboptimal solutions. Nevertheless, this procedure runs in polynomial time, as opposed to integer programming. It is important to note, however, that solutions to convex relaxations of combinatorial problems need not be integral, and in this case naive rounding techniques can lead to violations of the network flow constraints. We obtain integer solutions by showing that our convex relaxations are totally unimodular linear programs. A linear program being totally unimodular means that it always has optimal solutions that are integer valued  \cite{AhujaMagnantiEtAl1993}, and can thus be obtained using standard interior point optimization methods.

Network flow minimization problems are linear programs with constraints of the form \eqref{eq:flow-conservation}, and preserve total unimodularity. However, in the case of problem \eqref{monte_carlo_bundled}, the inclusion of the constraints \eqref{eq:waiter-conservation} break this totally unimodular structure, and hence solving a relaxation of \eqref{monte_carlo_bundled} with the $x,w \in \mathbb{N}^{n^2T}$ constraint removed is not guaranteed to return an integer solution. Alternatively, if we first assign vehicles to service existing customers, then the problem of rebalancing the empty vehicles no longer has constraints of type \eqref{eq:waiter-conservation}, and becomes totally unimodular. Inspired by this fact, in section \ref{subsec: bp_match} we discuss a bipartite matching algorithm we use to assign vacant vehicles to waiting customers, and in Section~\ref{subsec: tum_reb_flow} we solve a totally unimodular version of \eqref{monte_carlo_bundled} to determine rebalancing tasks.

\subsection{Bipartite Matching for Servicing Waiting Customers}\label{subsec: bp_match}
\indent We use a bipartite matching algorithm to pick up waiting customers in a way that minimizes the total waiting time. Specifically, the current state of the system is $z,d \in \mathbb{N}^n$, where $z_i$ is the number of vehicles currently available at station $i$, and $d_i$ is the number of outstanding customers at station $i$. The decision variable is a vector $x \in \mathbb{R}^{n^2}$ where $x_{n i + j}$ is the number of vehicles sent from station $i$ to station $j$. Let $A := \mathds{1}_n^T \otimes I_n - I_n \otimes \mathds{1}_n^T$, where $\mathds{1}_n$ is the vector of all $1$'s in $\mathbb{R}^d$, $I_n$ is the identity matrix of size $n \times n$ and $\otimes$ is the matrix Kronecker product. Using this notation, the resulting state of taking action $x$ in vehicle state $z$ is simply $z + Ax$. To satisfy the customers, we want $Ax + z \geq y$ elementwise. If this is not possible, we will pay a cost of $c_\lambda$ for every customer we do not pick up. To capture this, we define a drop vector $u = (y - Ax -z)_+$. The cost vector $c \in \mathbb{R}^{n^2}$ is defined so that $c_{i \cdot n + j}$ is the travel time between $i,j$. Thus, the optimal solution to the bipartite matching problem is obtained by solving the following linear program:

{\small
\begin{align}\label{bp_match}
\underset{x,u}{\text{min. }} & c^T x + \mathds{1}_n^T u\\
\text{ s.t. } & u \succeq 0 \nonumber \\
&u \succeq y - (Ax + z) \nonumber \\
&x \in \mathbb{R}^{n^2}, u \in \mathbb{R}^n \nonumber.
\end{align}
}

It can be shown that bipartite matching has the totally unimodular property, and, therefore, will return integer solutions when the constraints are also integer.

\subsection{Network Flow Optimization for Rebalancing Vehicles}\label{subsec: tum_reb_flow}

\indent To rebalance vacant vehicles in anticipation of future demand, we now solve \eqref{monte_carlo_bundled} with $w = 0$  to obtain a rebalancing flow. We have $w = 0$ because the task of picking up outstanding customers is given to a bipartite matching algorithm, and hence does not need to be considered here. In this case, we can relax the integer constraints on $x$ to obtain a totally unimodular linear program according to Lemma~\ref{TUM_relax}.\\ 
\begin{lemma}[Totally Unimodular SAA Rebalancing Problem]\label{TUM_relax}
\textit{Consider the following convex relaxation of \eqref{monte_carlo_bundled} where $w = 0$:
{\small 
\begin{align}\label{disjoint_LP}
\underset{x,w}{\text{min. }} & c_x^T x +\frac{1}{K} \sum_{k=1}^K  \sum_{ijt} u_{ijt}^k \\
\text{ s.t.} & \sum_{j = 1}^n  x_{ijt} - x_{ji(t - \tau_{ji})} = s_{it} \text{ for all} \nonumber \\
&i \in [n] \text{ and } t_0 < t \leq t_0 + T \nonumber \\
& u_{ijt}^k \geq 0 \; \; \; \; \; \; \; \; \forall k \in [K],i,j \in [n],t \in \mathcal{T}, \nonumber \\
& u_{ijt}^k \geq \lambda_{ijt}^k - x_{ijt} \; \; \; \; \forall k \in [K],i,j \in [n],t \in \mathcal{T} \nonumber \\
& \{u^k\}_{k=1}^K, x \in \mathbb{R}^{n^2 T} \; \; \forall k \in [K],i,j \in [n],t \in \mathcal{T} \nonumber.
\end{align}
}
This problem is totally unimodular.} \\
\end{lemma}
\noindent See section \ref{app: TUM_relax_proof} for a proof.

Thus, in the setting where $w = 0$, the convex relaxation from \eqref{monte_carlo_bundled} to \eqref{disjoint_LP} is tight in the sense that the solution to the latter is feasible and optimal for the former. For practical use the control strategy is to perform the tasks specified by the solutions to \eqref{bp_match} and \eqref{disjoint_LP}. The main strength of this approach is that both optimizers efficiently solve linear programs, as opposed to  integer programs like \eqref{monte_carlo_bundled} which can take orders of magnitude longer to solve in practice. 

\section{Stochastic optimization for Model Predictive Control of AMoD systems}\label{sec: MPC}
When controlling an autonomous fleet of cars in real time, using a receding horizon framework allows the controller to take advantage of new information that is observed in the system. We propose a model predictive control approach to control AMoD systems online whereby a controller periodically issues commands obtained from solutions to optimization problems. Algorithm \ref{alg:online_control} outlines the details of the MPC controller for one timestep. Every $\Delta t$ minutes, the controller queries the system to obtain information about the current state $\{s_{it}\}_{i \in \mathcal{V}, t\in \mathcal{T}}$, the current number of waiting customers $\lambda_0$, and recent demand measurements $\rho$. The controller then draws $K$ samples from $\widehat{P}(\lambda | \rho)$ and uses those samples to solve a stochastic optimization problem. The solve mode $\mathcal{I}$ specifies if a solution results from integer programming (cf. \ref{subsec: SAA_prob}) or linear programming (cf section \ref{sec: convex-relax}). Specifically, if $\mathcal{I} = 1$, the controller solves the integer program specified by \eqref{monte_carlo_bundled}, otherwise it solves the convex relaxation specified by \eqref{bp_match} and \eqref{disjoint_LP}. The controller executes the plan resulting from the optimization for the next $\Delta t$ seconds after which it repeats this process with updated information. 
\begin{algorithm}
    \SetKwInOut{Input}{Input}
    \SetKwInOut{Output}{Output}

    \underline{Stochastic AMoD Control} $(\mathcal{I}, s, \lambda_0, \rho)$\;
    Parameters: Road Network $G = (\mathcal{V}, \mathcal{E})$, Conditional generative demand model $\widehat{P}$\;
    \Input{Solve mode $\mathcal{I}$, System state $\{s_{it}\}_{i \in \mathcal{V}, t \in \mathcal{T}}$, Waiting customers $\lambda_0$, recent demand $\rho$.}
    \Output{Control action $x$.}
    Sample $\{\lambda^k\}_{k=1}^K \overset{\text{i.i.d.}}{\sim} \widehat{P}(\lambda | \rho)$\;
	\eIf
	{
	$\mathcal{I} = 1$\;
	}
	{
	Obtain $\{x_{\text{saa}}(t)\}_{t \in \mathcal{T}}$ by solving \eqref{monte_carlo_bundled} with samples $\{\lambda^k\}_{k=1}^K $\;
	\textbf{return} $x_{\text{saa}}(1)$ \;	
	}
	{
	Obtain $\{x_{\text{bm}}(t)\}_{t \in \mathcal{T}}$ by solving \eqref{bp_match} for waiting customers $\lambda_0$\;	
	Obtain $\{x_{\text{saa}}(t)\}_{t \in \mathcal{T}}$ by solving \eqref{disjoint_LP} with samples $\{\lambda^k\}_{k=1}^K $\;
	\textbf{return} $x_{\text{bm}}(1), x_{\text{saa}}(1)$. 	
	}
	\caption{Model Predictive Control for AMoD systems using Stochastic Optimization}\label{alg:online_control}
\end{algorithm}

\section{Numerical Experiments}\label{sec:numerical-experiments}
In this section we evaluate the performance of Algorithm \ref{alg:online_control} in a MPC framework. We simulate the operation of an AMoD system servicing requests from the two different datasets and compare performance to recent state of the art algorithms. The AMoD system services trip requests in Hangzhou, China from a DiDi Chuxing ridesharing company dataset in the first experiment, and requests from the New York City Taxi and Limousine Commission dataset in the second experiment. 

\subsection{Scenarios}
For Hangzhou, we leveraged a dataset provided by the Chinese ridesharing company Didi Chuxing. The dataset contains all trips requested by users from January 1 to January 21, 2016, resulting in a total of around eight million trips. The dataset separates Hangzhou into 793 discretized regions. However, the dataset contains only trips that started in a core subset consisting of 66 regions. For simplicity, we disregard trips that do not start and end in this core subset (approximately one million trips). For each trip, the records provide origin region, destination region, a unique customer ID, a unique driver ID, the start timestamp and the price paid. The dataset contains neither geographic information about the location of the individual districts, nor information on the duration of the trip. Thus, we used RideGuru \cite{ios_RideGuru:2017} to estimate the travel time of each trip from the trip price, which in turn allowed us to infer average travel times between regions. For the simulation, we used the first 15 days to train the forecasting model, and the last day to test in simulation by ``playing back'' the historical demand.

The second scenario is based on the well-known New York City Taxi and Limousine Commission dataset\footnote{\url{http://www.nyc.gov/html/tlc/html/about/trip_record_data.shtml}}. It contains, among others, all yellow cab taxi trips from 2009 to 2018 in New York City. For each trip, the start and end coordinates and timestamps are provided. For our simulation, we looked only into the trips that started and ended in Manhattan. Additionally, we partitioned the island into 50 regions. We used the trips between December 22, 2011 and February 29, 2012 to train the forecasting model, and used the evening rush hour (18:00-20:00) of March 1, 2012 for testing in simulation.

\subsection{Experimental Design}
For each scenario, we simulate the operation of an AMoD system by ``playing back'' the historical demand at a 6 second resolution. That is, vehicle and customer states get updated in 6 second timesteps. If on arrival, a customer arrives to a region where there is an available vehicle, the customer is assigned to that vehicle. Otherwise, the customer joins the region's customer queue. A customer's trip duration corresponds to the travel time recorded in the dataset. However, vehicle speeds are such that travel time between any two region centroids corresponds to the average travel time between those respective regions in the dataset.

Every $\Delta t = 5$ minutes, the simulation invokes an AMoD controller. The controller returns the rebalancing tasks for each region. These tasks, in turn, are assigned to idle vehicles as they become available. After $\Delta t$ minutes, unused tasks are discarded, and the controller is invoked again. We tested the following controllers:
\begin{itemize}
	\item \textit{Reactive} is a time-invariant reactive controller presented in \cite{PavoneSmithEtAl2012} which rebalances vehicles in order to track uniform vehicle availability at all stations.
	\item \textit{MPC-LSTM-MILP} is the model predictive controller presented in \cite{IglesiasRossiEtAl2018} which relies on point forecasts and mixed integer linear programming.
	\item \textit{MPC-LSTM-LP} is a relaxation of the MPC-LSTM-MILP controller attained by the ideas described in Section \ref{sec: convex-relax} by running two linear programs.
	\item \textit{MPC-LSTM-SAA} is the controller implementing Algorithm \ref{alg:online_control} with $\mathcal{I} = 0$ and $K = 100$ samples.
	\item \textit{MPC-Perfect} is a non-causal golden standard where the MPC controller is given perfect forecasts instead of samples of predicted demand. 
\end{itemize}
All MPC controllers are using a planning horizon of 4 hours.

\subsection{Forecasting}
In these experiments, the generative model $\widehat{P}$ for Algorithm \ref{alg:online_control} first estimates the mean of the future demand using a Long Short Term Memory (LSTM) neural network. The LSTM networks were trained on a subset of the data that does not include the test day. We trained a different network for each of the scenarios. Specifically, the LSTM takes as input the past 4 hours of observed customer demand and then predicts the expected demand for the next 2 hours. We assume that the demand follows a Poisson distribution. Moreover, to account for model uncertainty, we sample from the LSTM with dropout, a standard procedure to approximate Bayesian inference \cite{GalGhahramani2016}. Thus, we draw $K$ samples $\{\overline{\lambda}^k\}_{k=1}^K$ from the LSTM using dropout, and then sample the demand predictions from a Poisson process whose mean is specified by $\overline{\lambda}$ so that $\lambda^k \sim \text{Poisson}(\overline{\lambda}^k)$

\subsection{Results}

\begin{figure*}
	\centering
	\includegraphics[height=0.2\textheight]{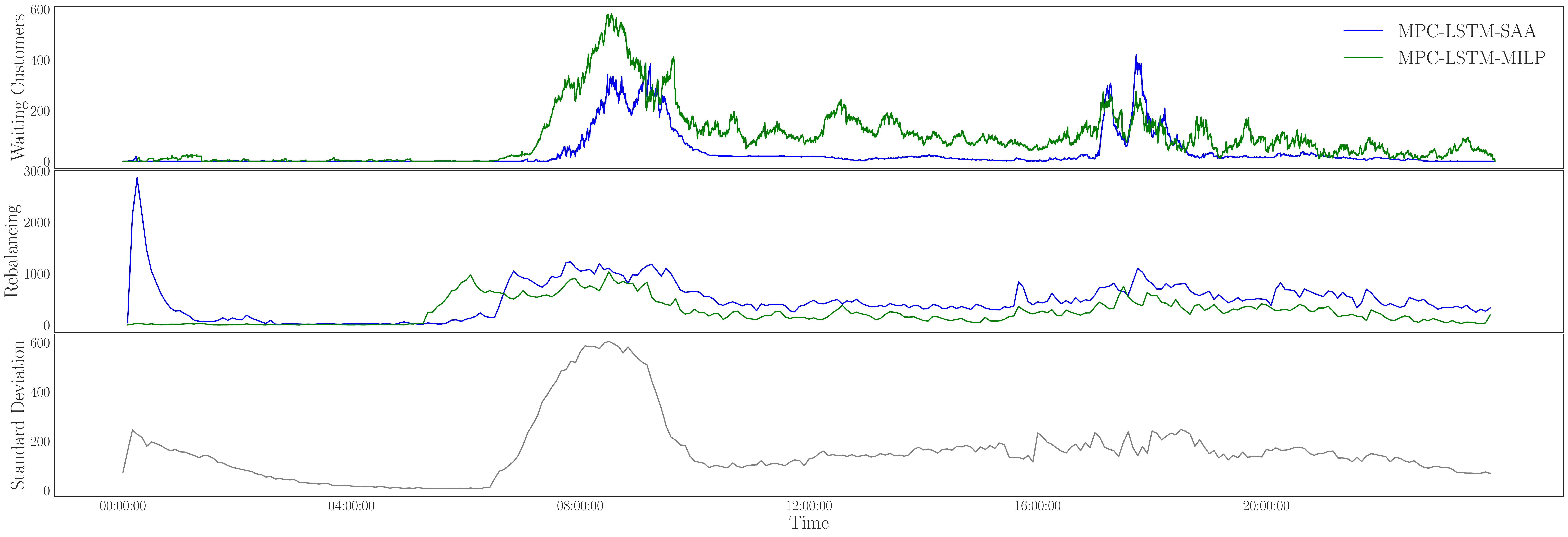}
	\caption{ The top plot shows the number of waiting customers as a function of time for both our controller (MPC-LSTM-SAA) in blue and the controller from \cite{IglesiasRossiEtAl2018} in green. Under the operation of our controller, there are significantly fewer waiting customers throughout the day, which is reflected by the waiting times. Unexpectedly, as a price of this improved service, our controller issues more rebalancing requests, since it is planning for many outcomes, as shown by the middle plot. Looking at the middle and bottom plots together, we see that our controller does additional rebalancing precisely when there is significant variance in the demand. }\label{fig: didi_exps}
\end{figure*}

In the Hangzhou scenario, the MPC-LSTM-SAA controller, based on \ref{alg:online_control}, greatly outperforms the other controllers: it provided a 62.3\% reduction in mean customer waiting time over the MPC-LSTM-MILP controller from \cite{IglesiasRossiEtAl2018}, and a 96.7\% reduction from Reactive (see Table \ref{tab: didi_summary}). Qualitatively, Figure \ref{fig: didi_exps} shows how the MPC-LSTM-SAA controller shows the greatest improvement over MPC-LSTM-MILP in times of the day where there is relatively high variance (in day-to-day travel demand variation). This suggests that the proposed algorithm's rebalancing strategy is better at handling future demand with high uncertainty than prior work. Naturally, handling uncertainty requires being prepared for a large variety of demand realizations. Thus, it is not unexpected that, as seen in Figure \ref{fig: didi_exps} and Table \ref{tab: didi_summary}, MPC-LSTM-SAA rebalances slightly more than MPC-LSTM-MILP; nonetheless, it still issues less rebalancing tasks than Reactive. 

Moreover, the performance of the MPC-LSTM-MILP and MPC-LSTM-LP controllers are essentially the same, which suggests that the relaxations described in \ref{sec: convex-relax} yield reliable runtimes without significantly sacrificing performance quality.  

\begin{table}[ht]
	\caption{Wait times for the DiDi scenario (seconds).}\label{tab: didi_summary}
	\centering
	\begin{tabular}{|r|cccc|}
	\hline 
	Wait Times: & Mean & Median & 99 Percentile & Reb. Tasks\\
	\hline
	Reactive & 276.284 & 72.0 & 1890.0 & 139927\\
	MPC-LSTM-MILP 	& 24.149 & \textbf{0.0} & 582.0 & 40097\\
	MPC-LSTM-LP 	& 23.305 & \textbf{0.0} & 558.0 & \textbf{39687}\\
	MPC-LSTM-SAA 	& \textbf{9.0799} & \textbf{0.0} & \textbf{264.0} & 68150\\
	MPC-Perfect 	& 5.527 & 0.0 & 168.0 & 32950 \\
	\hline
	\end{tabular}
\end{table}

The New York City scenario also demonstrates benefits of using stochastic optimization in the control. Table \ref{tab: nyc_summary} summarizes the results of this case study. While the 99 percentile wait time for the deterministic algorithm MPC-LSTM-LP is 32 percent smaller than that of Reactive, its mean waiting time is larger by 16 percent. Leveraging stochastic optimization, MPC-LSTM-SAA further improves the 99 percentile wait time of MPC-LSTM-LP by 17 percent and offers a 22 percent reduction in mean waiting time over Reactive. In summary, MPC-LSTM-SAA outperforms both Reactive and MPC-LSTM-LP in both mean and 99 percentile wait times. As a tradeoff, both MPC-LSTM-LP and MPC-LSTM-SAA issue more rebalancing tasks than Reactive. 

\begin{figure*}
	\centering
	\includegraphics[height=0.2\textheight]{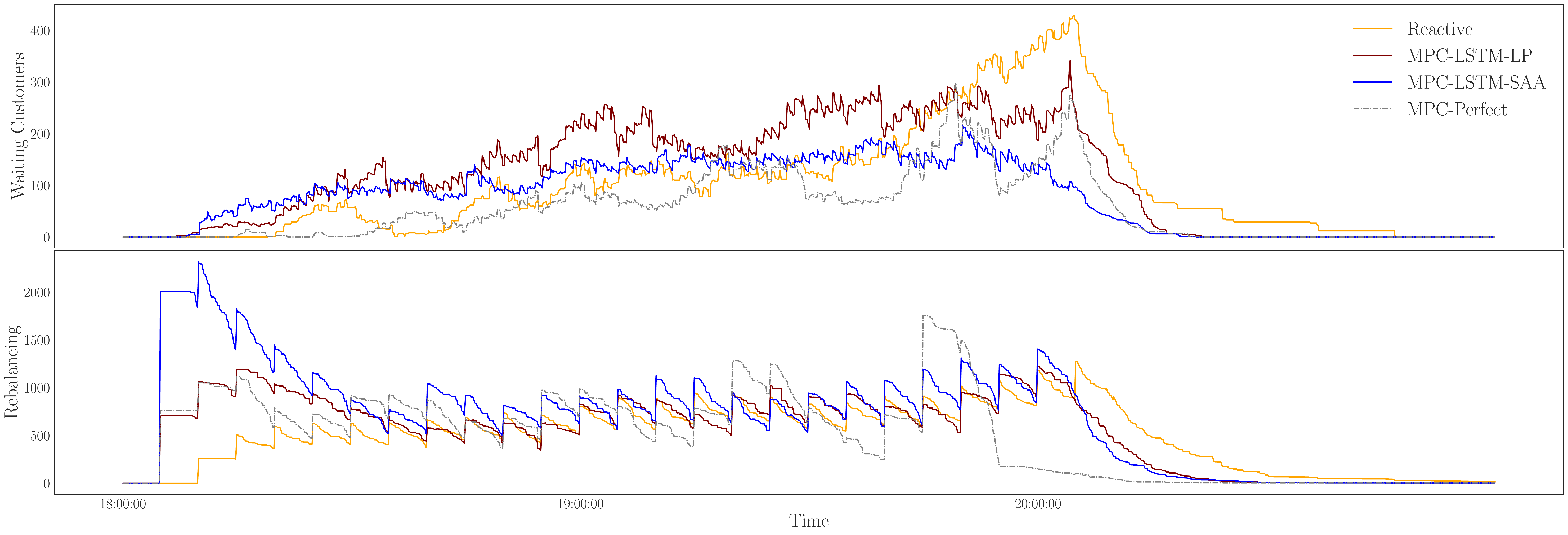}
	\caption{The top plot shows the number of waiting customers as a function of time for several controllers. As expected, compared to the reactive controller, the predictive controllers have more customers wait at the beginning of the simulation in order to better prepare for the customers appearing later. Leveraging stochastic optimization, MPC-LSTM-SAA outperforms MPC-LSTM-LP and Reactive in terms of mean waiting time. As a tradeoff, the bottom plot shows that the reactive controller issues the least amount of rebalancing tasks, while MPC-LSTM-SAA issues the most. }\label{fig: nyc_exps}
\end{figure*}

\begin{table}[ht]
	\caption{Wait times for the NYC scenario. (seconds)}\label{tab: nyc_summary}
	\centering
	\begin{tabular}{|r|cccc|}
	\hline 
	Wait Times: & Mean & Median & 99 Percentile & Reb. Tasks\\
	\hline
	Reactive & 19.15 & \textbf{0.0}  & 732.0 & \textbf{7196}\\
	MPC-LSTM-LP 	& 22.70 & \textbf{0.0} & 504.0 & 7907\\
	MPC-LSTM-SAA 	& \textbf{15.07} & \textbf{0.0} & \textbf{420.0} & 10952\\
	MPC-Perfect 	& 10.8 & 0.0 & 384.0 & 8356 \\
	\hline
	\end{tabular}
\end{table}

\section{Conclusions}\label{sec:conclusions}
\indent In this paper, we developed a stochastic Model Predictive Control algorithm for AMoD systems that leverages uncertain travel demand forecasts. We discussed two variants of the proposed algorithm, one using integer programming and a relaxed, linear programming approach that trades optimality for scalability. Through experiments, we show that the latter algorithm outperforms state-of-the art approaches in the presence of uncertainty. 

Future work will incorporate traffic congestion by modeling a detailed road network with finite capacities as is done in \cite{ZhangRossiEtAl2016}. This, in turn, can be coupled with models for public transit to provide a multi-modal \cite{SalazarRossiEtAl2018}, real-time stochastic control of AMoD systems. Similarly, ongoing research is studying coordination between the power network and electric AMoD systems \cite{Rossi2018}. Such a setting would be particularly interesting since both travel and power demand are stochastic. Due to its central role in the proposed algorithm, another area of interest is the development of principled algorithms for predicting short-term demand. In particular, we will tackle the challenge of capturing spatiotemporal demand distribution in the face of a myriad of factors, such as weather, traffic and vehicle availability. Finally, accurate forecasts will enable robust and risk-averse objectives.

\bibliographystyle{IEEEtran}
\bibliography{../../../bib/main,../../../bib/ASL_papers}

\section{Appendix}

\subsection{Proof of Lemma \ref{triangle}}\label{app: triangle_proof}
Given $f,g : \mathcal{X} \rightarrow \mathbb{R}$ define $x_f := \arg\min_{x \in \mathcal{X}} f(x)$ and $x_g := \arg\min_{x \in \mathcal{X}} g(x)$. Let $||f-g||_\infty := \sup_{x \in X} |f(x) - g(x)|$. We then see that
\begin{align*}
f(x_g) &\leq g(x_g) + ||f-g||_\infty \\
&\leq g(x_f) + ||f-g||_\infty \\
&\leq \left[ f(x_f) + ||f-g||_\infty \right] + ||f-g||_\infty \\
&= f(x_f) + 2 ||f-g||_\infty,
\end{align*}
which is the desired result. 

\subsection{Proof of Lemma \ref{bound}}\label{app: bound_proof}
\noindent First recall the definition of the $\chi^2$-divergence $\chi^2(P||Q)$ between two probability distributions $P,Q$.
{\small
\begin{align*}
\chi^2(P||Q) &:= \mathbb{E}_Q \left[ \left(1 - \frac{dP}{dQ} \right)^2 \right] \text{ if } P \ll Q, +\infty \text{ else.}
\end{align*}
}
The function $\chi^2(P||Q)$ is non-negative and is zero if and only if $P = Q$. Since $(\cdot)_+$ is $1$-Lipschitz, we have: 

{\small
\begin{align*}
&F(x,w) - \mathbb{E}_{\widehat{P}}\widehat{F}_K(x,w) \\
&=\sum_{ijt} \sum_{\lambda \in \mathbb{N}} (\lambda + w_{ijt} - x_{ijt})_+ (P_{ijt}(\lambda) - \widehat{P}_{ijt}(\lambda)) \\
\end{align*}
Define $\ell_{ijt}(\lambda) := (\lambda + w_{ijt} - x_{ijt})_+ - \mathbb{E}_{P_{ijt}} (\lambda + w_{ijt} - x_{ijt})_+$, we have:
\begin{align*}
&F(x,w) - \mathbb{E}_{\widehat{P}}\widehat{F}_K(x,w) = \sum_{ijt} \sum_{\lambda \in \mathbb{N}} \ell_{ijt}(\lambda) (P_{ijt}(\lambda) - \widehat{P}_{ijt}(\lambda))
\end{align*}
Since $\sum_{\lambda \in \mathbb{N}} C (P_{ijt}(\lambda) - \widehat{P}_{ijt}(\lambda)) = 0$ for any constant $C$, we let $C = \mathbb{E}_{P_{ijt}} (\lambda + w_{ijt} - x_{ijt})_+$.
\begin{align*}
&\sum_{ijt} \sum_{\lambda \in \mathbb{N}} \ell_{ijt}(\lambda) (P_{ijt}(\lambda) - \widehat{P}_{ijt}(\lambda)) \\
&= \sum_{ijt} \sum_{\lambda \in \mathbb{N}} \ell_{ijt}(\lambda) \sqrt{P(\lambda)_{ijt}} \left( \sqrt{P(\lambda)_{ijt}} - \frac{\widehat{P}_{ijt}(\lambda)}{\sqrt{P_{ijt}(\lambda)}} \right) \\
&\leq \sqrt{ \sum_{\substack{ijt \\ \lambda \in \mathbb{N}}} \ell_{ijt}(\lambda)^2 P_{ijt}(\lambda) } \sqrt{ \sum_{\substack{ijt \\ \lambda \in \mathbb{N}}}  \left( 1 - \frac{\widehat{P}_{ijt}(\lambda)}{P_{ijt}(\lambda)} \right)^2 } \\
&= \sqrt{\text{Var}_P[||(\lambda + w - x)_+||_2}  \sqrt{\sum_{ijt} \chi^2(\widehat{P}_{ijt}|| P_{ijt})} \\
&\leq \sqrt{\text{Var}_P(||\lambda||_2)} \; ||\chi(\widehat{P}|| P)||_2
\end{align*}
}
The first inequality is due to Cauchy-Schwarz in $L^2_{P}$, and the second inequality is due the fact that $\text{Var}(X_+) \leq \text{Var}(X)$ for any random variable $X$ by the following calculations:
\begin{align*}
\text{Var}(X_+) &:= \mathbb{E}[(X_+ - \mathbb{E}[X_+])^2] \\
&\leq \mathbb{E}[(X_+ - \mathbb{E}[X_+])^2] + (\mathbb{E}[X_+] - \mathbb{E}[X]_+)^2 \\
&= \mathbb{E}[(X_+ - \mathbb{E}[X]_+)^2] \\
&\leq \mathbb{E}[(X - \mathbb{E}[X])^2] \\
&= \text{Var}(X)
\end{align*} 
The second inequality is because $(\cdot)_+$ is a 1-Lipschitz function. It is also possible to control the model error using the RMSE of the generative model $\widehat{P}$, but that bound is weaker than what is presented here. For the standard deviation bound, we use the concentration of measure for sub-exponential random variables. A random variable $X$ is sub-exponential if there exists parameters $\sigma^2, b$ so that for any $|\lambda| \leq b^{-1}$, we have

{\small 
\begin{align*}
\log \mathbb{E} [e^{\lambda(X - \mathbb{E} X)}] &\leq \frac{\lambda^2 \sigma^2}{2}
\end{align*}
}
The following probability bounds for sub-exponential random variables are well known: If $X$ is $(\sigma^2, b)$ sub-exponential, then:
{\small
\begin{align*}
\mathbb{P}[|X - \mathbb{E}X| > t] &\leq \exp \left( -\frac{t^2}{2\sigma^2} \right) \text{ if } t \leq \frac{\sigma^2}{b} \\
\mathbb{P}[|X - \mathbb{E}X| > t] &\leq \exp \left(- \frac{t}{2b} \right) \text{ otherwise}
\end{align*}
}
Let $\{\lambda^1, ... , \lambda^K\}$ be samples from $\widehat{P}$ used to form the objective function. Let $\widehat{F}_{x,w}(\lambda) := \sum_{ijt} (\lambda_{ijt} + w_{ijt} - x_{ijt})_+$. 1-Lipschitz functions of sub-exponential random variables are also sub-exponential with the same parameters, thus $\{\widehat{F}_{x,w}(\lambda^k)\}_{k=1}^K$ are i.i.d. $(\sigma^2, b)$-sub-exponential random variables. Thus, the objective of \eqref{monte_carlo_bundled}, $\frac{1}{K} \sum_{k=1}^K \widehat{F}_{x,w} (\lambda^k)$ is $(\frac{\sigma^2}{K}, \frac{b}{K})$ sub-exponential. Applying the first part of the probability bound, we see that:
{\small
\begin{align*}
\mathbb{P} \left( \left| \frac{1}{K} \sum_{k=1}^K \widehat{F}_{x,w}(\lambda^k) - \mathbb{E} \widehat{F}_{x,w}(\lambda) \right| > t \right) &\leq \exp \left( - \frac{K t^2}{2 \sigma^2} \right)
\end{align*}
}
for any $t < \frac{\sigma^2/K}{b/K} = \sigma^2/b$. For any $\delta > 0$ error tolerance, setting $t = \frac{2 \sigma}{\sqrt{K}} \sqrt{n^2 T \log (m) + \log (\delta^{-1/2})}$, for sufficiently large $K$ the bound evaluates to $\delta (m)^{-n^2 T}$. However this inequality only applies to a particular pair of $x,w$. Since $x \in \mathbb{R}^{n^2T}$ and $||x||_\infty \leq m$, $x$ can take at most $|m|^{n^2 T}$ many distinct values. Note that if $w_{ijt} > m$ we can always set $w_{ijt} = m$ without affecting performance because the system cannot pick up more than $m$ waiting customers at any time. Thus, we also have $||w||_\infty \leq m$ and hence $w$ can take at most $m^{n^2 T}$ values. Thus there are at most $m^{2 n^2 T}$ possible plans $(x,w)$. Taking a union bound over all possible $x,w$ gives, with probability at least $1-\delta$,

{\small
\begin{align*}
&\left| \left |\frac{1}{K} \sum_{k=1}^K \widehat{F}_{x,w}(\lambda^k) - \mathbb{E} \widehat{F}_{x,w} (\lambda) \right| \right|_\infty \\
&\leq \frac{2 \sigma}{\sqrt{K}} \sqrt{n^2 T \log (m) + \log (\delta^{-1/2})}
\end{align*}
}

Applying lemma \ref{triangle} with this bound yields the desired result. 

\subsection{Proof of Lemma \ref{num_vars}}\label{app: num_vars_proof}
Here we use sub-exponential concentration inequalities to obtain a bound on the maximum and minima of i.i.d. sub-exponential random variables. Lemma \ref{num_vars} is related to the the distribution of maxima of sub-exponential random variables. Let $X_1,...,X_n$ be i.i.d. zero mean $(\sigma^2,b)$ sub-exponential random variables. We proceed by the standard Chernoff bounding technique. For any $0 < \lambda \leq b^{-1}$, we have:

{\small
\begin{align*}
\mathbb{P} \left[ \max_{1\leq i \leq n} X_i \geq t \right] &= \mathbb{P} \left[ e^{\lambda (\max_{1\leq i \leq n} X_i )}\geq e^{\lambda t} \right] \\
&= \mathbb{P} \left[ \max_{1 \leq i \leq n} e^{\lambda X_i} \geq e^{\lambda t} \right] \\
\text{Markov Inequality} \rightarrow &\leq e^{-\lambda t} \mathbb{E} \left[ \max_{1 \leq i \leq n} e^{ \lambda X_i} \right]  \\
&\leq e^{-\lambda t} \mathbb{E} \left[ \sum_{1 \leq i \leq n} e^{ \lambda X_i} \right] \\
&\leq n \exp \left( - \lambda t \right) \exp \left( \frac{\lambda^2 \sigma^2}{2} \right) \\
&= \exp \left( - \lambda t + \frac{\lambda^2 \sigma^2}{2} + \log n\right)
\end{align*}
}
If $t > \frac{\sigma^2}{b}$ then setting $\lambda = \frac{1}{b}$ gives the tightest upper bound, in which case we have:
{\small
\begin{align*}
\mathbb{P} \left[ \max_{1\leq i \leq n} X_i \leq t \right] &\leq \exp \left( - \frac{t}{b} + \frac{\sigma^2}{2b^2} + \log n \right)
\end{align*}
}
but recall that $t > \frac{\sigma^2}{b} \implies \frac{\sigma^2}{2b^2} \leq \frac{t}{2b}$, meaning 
{\small
\begin{align*}
\mathbb{P} \left[ \max_{1\leq i \leq n} X_i \leq t \right] &\leq \exp \left( -\frac{t}{2b} + \log n \right)
\end{align*}
}
thus for any $\delta > 0$, setting $t = 2b \log \frac{n}{\delta}$, the upper bound is equal to $\delta$. Hence,
{\small
\begin{align*}
\mathbb{P} \left[ \max_{1\leq i \leq n} X_i \geq 2b \log \frac{n}{\delta} \right] \leq \delta
\end{align*}
}
The concentration of the minimum is analogous, by noting that $-X$ is also sub-exponential, and applying the above argument. Applying this to our problem, if the demand $\lambda_{ijt}$ for each $(i,j,t)$ is $(\sigma^2_{ijt},b)$ sub-exponential, and $K$ samples $\lambda_{ijt}^1,..., \lambda_{ijt}^K$ are observed, then by the above argument, with probability at least $1-\frac{\delta}{n^2T}$ all samples fall in the interval  $$\left[\mathbb{E}[\lambda_{ijt}] - 2b \log \frac{K n^2 T }{\delta}, \mathbb{E}[\lambda_{ijt}] + 2b \log \frac{K n^2 T }{\delta} \right].$$ But since these samples are integer valued, if they lie in an interval of size $O(\log K)$, then there can be at most $O(\log K)$ distinct samples of the demand. Taking a union bound over all tuples $(i,j,t)$, we have that with probability at least $1-\delta$, for each tuple $(i,j,t)$, the number of unique elements in $\{\lambda_{ijt}^k\}_{k=1}^K$ is at most $4 b \log \frac{K n^2 T}{\delta}$. Summing over all $(i,j,t)$ we then have that the total number of decision variables is at most $4 b n^2 T \log \frac{K n^2 T}{\delta}$. The number of decision variables is trivially at most $K n^2 T $, therefore taking the better of the two bounds yields the result.

\subsection{Proof of Lemma \ref{TUM_relax}} \label{app: TUM_relax_proof}
Define $u = [u^1, ..., u^k] \in \mathbb{R}^{n^2 T K}_+$ so that the decision variable for \eqref{disjoint_LP} is $z = [x, u]$. To show that \eqref{disjoint_LP} is totally unimodular, it is necessary and sufficient to show that all extreme points of the constraint polyhedron are integer vectors. Recall that a point $z^* = [x^*, \; u^*]$ is an extreme point if and only if the matrix of active constraints $B(z^*)$ has rank $n^2 T(K+1)$, where the active constraint matrix is the matrix whose rows are the equality constraints and active inequality constraints of the problem at $z^*$. Since $z \in \mathbb{R}^{n^2T(K+1)}$, a point $z^*$ is extreme if and only if $B(z^*)$ has full column rank. We can express the active constraints as:
\begin{align*}
\left[ 
\begin{tabular}{cc}
$A$ & $0$ \\
$C$ & $D$
\end{tabular}
\right]
\left[ 
\begin{tabular}{cc}
$x^*$ \\
$u^*$
\end{tabular}
\right]
=
\left[ 
\begin{tabular}{cc}
$b$ \\
$0$
\end{tabular}
\right]
\end{align*}
where $A,b_1$ are chosen so that $Ax^* = b$ is equivalent to the network flow constraints specified by \eqref{eq:flow-conservation}, and $C,D$ are chosen so that $Cx^* + Du^* = 0$ represents the active inequality constraints $u_{ijt}^k = \lambda_{ijt}^k - x_{ijt}$, and/or $u_{ijt}^k = 0$. Noting that
\begin{align*}
B(z^*)z = 
\left[ 
\begin{tabular}{cc}
$A$ & $0$ \\
$C$ & $D$
\end{tabular}
\right]
\left[ 
\begin{tabular}{cc}
$x$ \\
$u$
\end{tabular}
\right]
=
\left[ 
\begin{tabular}{cc}
$Ax$ \\
$Cx + Du$
\end{tabular}
\right]
\end{align*}
$B(z^*)$ has full column rank only if $A$ has full column rank, meaning $x^*$ must be an extreme point of the polyhedral constraints defined by $A$. However, recalling that $A$ arises from network flow constraints and is a unimodular matrix, this immediately implies that $x^*$ must be an integer vector. For each tuple $(i,j,t,k)$, the decision variable $u_{ijt}^k$ is subject to exactly two constraints: $u_{ijt}^k \geq \lambda_{ijt}^k - x_{ijt}$, and $u_{ijt}^k \geq 0$. In any extreme point, at least one of these constraints is active. This can be shown easily via contradiction. If $z^*$ is extreme and for some $u_{ijt}^{k*}$, both (i.e. all) constraints are inactive, then define $\textbf{I}$ to be the index of $u_{ijt}^{k*}$ in $z^*$. Since there are no active constraints involving $u_{ijt}^{k*}$, the $\textbf{I}$th column of $B(z^*)$ is zero, hence $B(z^*)$ cannot have full column rank. Therefore $u_{ijt}^{k*} \in \{0,\lambda_{ijt}^k - x_{ijt} \}$. Since we showed $x^*$ is integer, and $\{\lambda^k\}_{k=1}^K$ are integer, this implies that $u^*$ must be integer, finally implying that $z^*$ is integer. Since all extreme points are integer valued, \eqref{disjoint_LP} is a totally unimodular linear program.

\end{document}